\documentclass[12pt]{article}
\usepackage{latexsym}
\usepackage{amssymb}
\usepackage{cite}
\usepackage{epsfig}
\title{\LARGE Chargino production via $Z^0$-Boson Decay in a
Strong Electromagnetic Field}
\author{\large A.V.Kurilin{\footnote{E-mail address:
kurilin@mail.ru}}\\ Moscow Technological Institute \footnote{on leave from Moscow State Open Pedagogical University},\\ Leninsky Prospect, 38a, Moscow, 119334, Russia\\}
\begin{document}
\begin{titlepage}
\maketitle \thispagestyle{empty}

\begin{abstract}{In the framework of MSSM the probability
of $Z^0$-boson decay to charginos in a strong electromagnetic
field, $Z^0\rightarrow \chi ^{+} \chi ^{-}$, is analyzed. The method of calculations employs exact solutions of
relativistic wave equations for charginos in a crossed electromagnetic field.
Analytic expression for the decay width
$\Gamma(Z^{0}\rightarrow \chi ^{+} \chi ^{-})$
is obtained at an arbitrary value of the parameter
$\varkappa=e m_Z^{-3}\sqrt{-(F_{\mu\nu}q^\nu)^2}$,
which characterizes the external-field strength $F_{\mu\nu}$ and $Z^0$-boson momentum $q^{\nu}$. The process $Z^0\rightarrow \chi ^{+} \chi ^{-}$ is forbidden in vacuum for the case of relatively heavy charginos: $M_{\chi^{\pm}} > m_Z/2$
. However in an intense electromagnetic background this reaction could take place in the region of superstrong fields ($\varkappa > 1$).}
\end{abstract}
\end{titlepage}

Minimal Supersymmetric extension of the Standard Model (MSSM) \cite{SUSY} is a neat solution of the famous hierarchy problem \cite{hierarchy} and it predicts a number of new particles to be discovered. In this large family of hypothetical "superparticles" there are charginos $\chi_1^\pm, \chi_2^\pm $ which arise as a mixture of winos $\tilde W^{\pm}$, the spin-$1/2$ superpartners of the gauge $W^{\pm}$ bosons, and higgsinos $\tilde H^{\pm}$, the fermion superpartners of the two scalar Higgs fields which break spontaneously the electroweak symmetry. Vacuum expectation values $v_1, v_2$ of the two Higgs fields can be characterized by the angle $\beta$ which is defined as usual by the ratio: $\tan\beta=v_2/v_1$. Chargino masses can be expressed in terms of the fundamental supersymmetry (SUSY) parameters $M_2$ and $\mu$:
\begin{equation}
\label{Mchi}
M^2_{\chi_1,\chi_2}=\frac{1}{2}
\left(\mu^2+M^2_2+2 m^2_W \mp \Delta \right),
\end{equation}
where $m_W$ is the $W$-boson mass and the quantity $\Delta$ determines the difference between the squares of chargino masses ($M_{\chi_2} > M_{\chi_1}$ ):
\begin{equation}
\label{Delta-chi}
\Delta=M_{\chi_2}^2-M_{\chi_1}^2=
\left[ \left(\mu^2+M^2_2+2 m^2_W \right)^2 -
4 \left(\mu M_2 - m^2_W \sin 2\beta \right)^2   \right]^{1/2}.
\end{equation}
For the sake of simplicity in present calculations it is assumed that the Higgs mass parameter $\mu$ and the $SU(2)$ gaugino masses $M_2$ arising in the Lagrangian of MSSM with other soft SUSY-breaking terms are real.

In the leading order of perturbation theory the matrix element
of $Z^0$-boson decay to a pair of the lighter charginos, $\chi_1^{\pm}$, is given by
\begin{equation}
\label{SZ_chi} S_{fi}=i\int d^4 x
\overline{\chi}_{1}(x,p)\gamma^\mu (g_{V1}+\gamma^5
g_{A1})\chi_{1 }^c(x,p') Z_\mu(x,q),
\end{equation}
where the vertex of $Z^0$-boson and $\chi_1^{\pm}$-chargino couplings can be described by the two constants:
\begin{equation}
\label{gv1}
g_{V1}=\frac{g}{8 \cos\theta_{\rm W}}
\left(2-8\cos^2\theta_{\rm W}-\cos 2\phi_R-\cos 2\phi_L \right)
\end{equation}
\begin{equation}
\label{gA1}
g_{A1}=\frac{g}{8 \cos\theta_{\rm W}}
\left(\cos 2\phi_R-\cos 2\phi_L \right)
\end{equation}
The two angles $\phi_R$ and $\phi_L$ define the relationship between charginos, winos and higgsinos and can be computed from the formulae:
\begin{equation}
\label{cos2fR}
\cos 2\phi_R=\frac{\mu^2-M^2_2-2 m^2_W \cos 2\beta}{\Delta},
\end{equation}
\begin{equation}
\label{cos2fL}
\cos 2\phi_L=\frac{\mu^2-M^2_2+2 m^2_W \cos 2\beta}{\Delta},
\end{equation}
whereas the Weinberg angle $\theta_{\rm W}$ determines the usual relationship between $W$-boson and $Z$-boson masses: $\cos\theta_{\rm W}=m_W/m_Z$.

\begin{figure}[t]

\setlength{\unitlength}{1cm}
\begin{center}
%\leavevmode
\epsfxsize=15.cm \epsffile{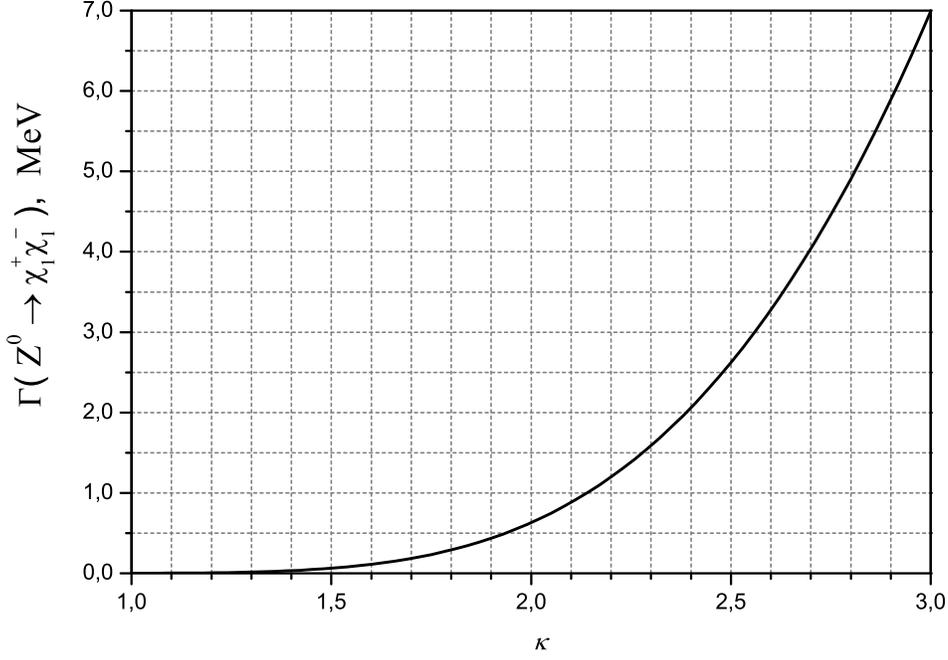}
\begin{minipage}[t]{15 cm}
\caption[]{Partial decay width of $Z^0$-boson into lighter charginos at relatively weak fields plotted as a function of parameter $\varkappa$ (\ref{invariantk}).}
\label{Gamma_Chi-1}
\end{minipage}
\end{center}
\end{figure}

The method of calculations in this paper is based
on the crossed-field model, which was successfully
applied in our previous investigations dealing with $W^{\pm}$ and
$Z^0$-bosons decays \cite{Kurilin-2004, Kurilin-2009} and with
SUSY processes in background electromagnetic fields \cite{Kurilin-1990,Kurilin-Ternov}.
The main idea of this approach is to describe interactions
with an external electromagnetic background by choosing specific wave
functions $\chi^-(x,p)$ and $\chi^+(x,p')$ for charginos
$\chi^{\pm}$ which are exact solutions of the Dirac equation
in a crossed electromagnetic field. This method makes it possible to
take into account the non-perturbative interactions with
the electromagnetic background and to obtain results for new reactions which are forbidden in vacuum (for a review see, for example, \cite{Borisov_UFN} and references therein). The processes of chargino production via $W^{\pm}$ and $Z^0$-bosons decays in vacuum  have been considered by many authors \cite{Chargino_Vacuum}. However present experimental limits on chargino masses \cite{PDG-2014} evidence that the reaction $Z^{0}\rightarrow \chi ^{+} \chi ^{-}$  is forbidden under usual conditions and	it inspires us to study the impact of strong electromagnetic fields on the decays mentioned above.
The wave functions for charginos in a background  electromagnetic field have the  simplest form for a crossed-field configuration, in which case the field-strength tensor $F_{\mu\nu}$ obeys the conditions:
\begin{equation}
\label{crossed}F_{\mu\nu} F^{\mu\nu}=F_{\mu\nu}\tilde
F^{\mu\nu}=0.
\end{equation}
The explicit form of theses wave functions is rather cumbersome and it can be found, for example, in \cite{Kurilin-1999,Kurilin-2009}.
Substituting the chargino wave functions into expression (\ref{SZ_chi})
for the $S$-matrix element and performing integration of
$\mid S_{fi}\mid^2$ over the phase space we obtain the probability of $Z$-boson decay into a pair of the lighter charginos $\chi_1^{\pm}$.
\begin{eqnarray}
\label{Z-chi-1} P(Z^0\rightarrow\chi^+_1\chi^-_1)
=\frac{G_{\rm F} m_Z^4 \  c_1 (1+\delta_1)}{96\sqrt{2}\ \pi^2 q_0} \int\limits_0^1 du
\biggl\{ \biggl[1+(3\rho_1-1)\lambda_1\biggr]
\Phi_1(z_1)-\nonumber\\
-\frac{2\varkappa^{2/3}}{\left[u(1-u)\right]^{1/3}}
\biggl[1-2u+2u^2+(1-\rho_1) \lambda_1\biggr]
\Phi'(z_1) \biggr\}.
\end{eqnarray}
The probability of chargino production in a crossed electromagnetic field  is expressed in terms of the Airy functions $\Phi'(z)$ and $\Phi_1(z)$ which
have the well-known integral representations.
\begin{equation}
\Phi(z)=\int\limits_0^\infty \cos\left(zt+{t^3\over
3}\right)dt,
\qquad
 \Phi_1(z)=\int\limits_z^\infty
\Phi(t) dt ,
\qquad \Phi'(z)=\frac{d\Phi(z)}{dz} .
\end{equation}
These functions depend on the argument
\begin{equation}
\label{z1} z_1=\frac{\lambda_1 - u(1-u)}{[\varkappa u
(1-u)]^{2/3}},
\end{equation}
which characterizes the electromagnetic-field-strength $F^{\mu\nu}$ and $Z$-boson momentum $q_{\nu}$ through the following invariant parameter
\begin{equation}
\label{invariantk}\varkappa=\frac{e}{m_Z^3}
\sqrt{-(F^{\mu\nu}q_\nu)^2}.
\end{equation}
The other dimensionless parameters in expression (\ref{Z-chi-1}) $\lambda_1, \rho_1, \delta_1$
are associated with chargino masses (\ref{Mchi}) and the coupling constants (\ref{gv1}),(\ref{gA1})

\begin{equation}
\label{lambda-c-1} \lambda_1 =\left({ M_{\chi_1} \over m_Z} \right)^2,
\qquad
c_1=\left(\cos 2\phi_R+\cos 2\phi_L +8\cos^2\theta_{\rm W}-2\right)^2,
\end{equation}
\begin{equation}
\label{rho1}
\rho_1=\frac{1-\delta_1}{1+\delta_1},
\qquad
\delta_1=\left(\frac{\cos 2\phi_R-\cos 2\phi_L}
{\cos 2\phi_R+\cos 2\phi_L +8\cos^2\theta_{\rm W}-2}\right)^2.
\end{equation}
Let now consider asymptotic estimates of the $Z$-boson partial decay width
$\Gamma(Z^0\rightarrow\chi^+_1\chi^-_1)$ in a crossed electromagnetic field at various values of the parameter $\varkappa$ (\ref{invariantk}). In the domain of relatively weak fields the $Z^0$-boson decay width into a pair of lighter charginos can be described by the formula:
\begin{eqnarray}
\label{Zchi-weak} \Gamma(Z^0\rightarrow \chi_1^+ \chi_1^-)=
G_{\rm F} m_Z^3 \  \frac{ c_1 (1+\delta_1) \lambda_1
\left( 5\rho_1+1+8\lambda_1 (1-\rho_1) \right)}
{64\pi\sqrt{6}\left(4\lambda_1-1\right)\sqrt{8\lambda_1+1}}\times \nonumber\\
\times \  \varkappa\
\exp\left[-\frac{\left(4\lambda_1-1\right)^{3/2}}
{3\varkappa}\right].
\end{eqnarray}
We see that the decay rate is exponentially suppressed at weak fields which is typical for processes being forbidden in vacuum. For numerical calculations presented below we employ the following parameters:
$$\tan\beta=5, \quad M_2=200\ \mbox{GeV},\quad  \mu=250\ \mbox{GeV}.$$
This choice corresponds to chargino masses $M_{\chi_1}=158\ \mbox{GeV}$ and $M_{\chi_2}=301\ \mbox{GeV}$ which can be easily obtained from equations (\ref{Mchi}),(\ref{Delta-chi}). In the area of relatively small values of the field-strength parameter $\varkappa$ the partial decay width $\Gamma(Z^0\rightarrow\chi^+_1\chi^-_1)$ into lighter charginos grows monotonously reaching the value $\Gamma_{\chi_1}\simeq 7\ {\rm MeV}$ at $\varkappa =3$. Exact results of calculations are  displayed in Fig.\ref{Gamma_Chi-1}  for the region $\varkappa \le 3$ where the probability of $Z^0$-boson  decay into a pair of heavier charginos $Z^0\rightarrow\chi^+_2\chi^-_2$ is negligibly small because of the large mass differences between $\chi_1^\pm$ and $\chi_2^\pm$. However in strong fields the $Z^0$-boson decays into heavier charginos $\chi_2^\pm$ become sizable and should be taken into account.

The probability of decay $Z^0\rightarrow\chi^+_2\chi^-_2$ can be obtained from equation (\ref{Z-chi-1}) by formal substitutions: $\lambda_1 \to \lambda_2, c_1 \to c_2, \rho_1 \to \rho_2, \delta_1 \to \delta_2$, where

\begin{equation}
\label{lambda-c-2}
\lambda_2 =\left({ M_{\chi_2} \over m_Z} \right)^2,
\qquad
c_2=\left(\cos 2\phi_R+\cos 2\phi_L -8\cos^2\theta_{\rm W}+2\right)^2,
\end{equation}
\begin{equation}
\label{rho2}
\rho_2=\frac{1-\delta_2}{1+\delta_2},
\qquad
\delta_2=\left(\frac{\cos 2\phi_R-\cos 2\phi_L}
{\cos 2\phi_R+\cos 2\phi_L - 8\cos^2\theta_{\rm W}+2}\right)^2.
\end{equation}
In the domain $\varkappa \gg 10$ the partial decay widths of $Z^0$-boson into charginos $\chi_i^\pm, i=1,2$ can be estimated by the the following equation:
\begin{eqnarray}
\label{Zchi-strong} \Gamma(Z^0\rightarrow \chi_i^+ \chi_i^-)=G_{\rm F} m_Z^3\
\frac{c_i (1+\delta_i)}{96\pi\sqrt{2}}
\left[ \frac{15\Gamma^4(2/3)}{14\pi^2}(3\varkappa)^{2/3} + \frac{1}{3}+\right.
\nonumber \\ \left.+\frac{3\Gamma^4(1/3)}{110\pi^2}(3\varkappa)^{-2/3}+
\frac{9\Gamma^4(2/3)}{104\pi^2}(3\varkappa)^{-4/3} \right].
\end{eqnarray}
%%%%%%%%%%%%%%%%%%%%%%%%%% Fig1%%%%%%%%%%%%%%%%%%%%%%%%%%%%%%%%%%%
%%%%%%%%%%%%%%%%%%%%%%%%%%%%%%%%%%%%%%%%%%%%%%%%%%%%%%%%%%%%%%%%%%%%

\begin{figure}[t]
\setlength{\unitlength}{1cm}
\begin{center}
%\leavevmode
\epsfxsize=15.cm \epsffile{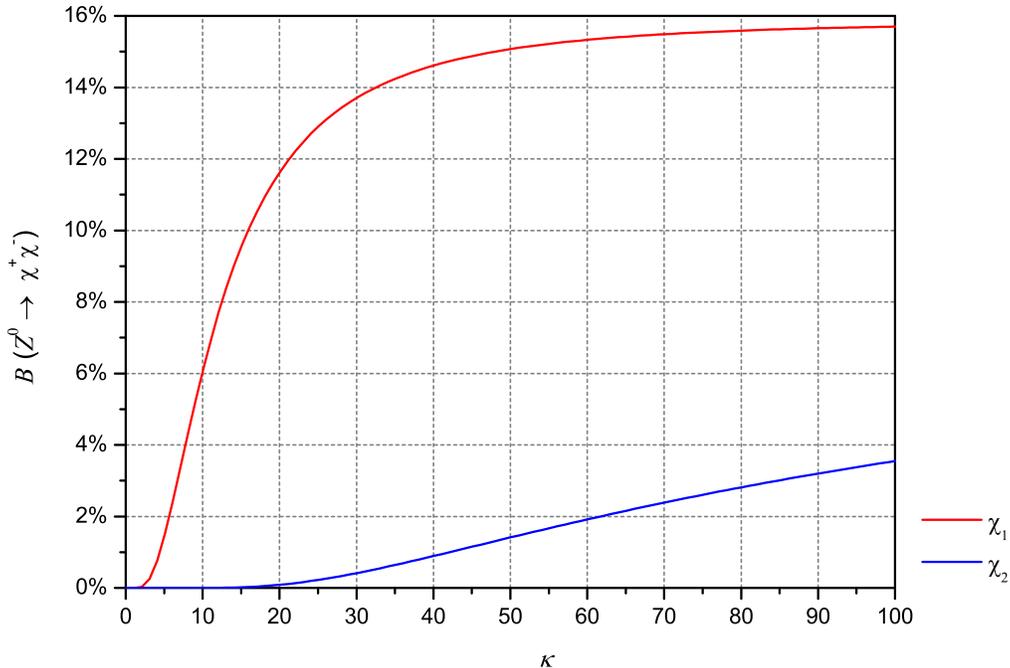}
\begin{minipage}[t]{15 cm}
\caption[]{Branching ratios $B(Z^0\rightarrow\chi^+_i\chi^-_i)= \Gamma(Z^0\rightarrow\chi^+_i\chi^-_i)/ \Gamma_{\rm tot}(\varkappa)$ of $Z$-boson decays into charginos $\chi_i^{\pm}, i=1,2$ in strong electromagnetic fields.}  \label{Branching-12}
\end{minipage}
\end{center}
\end{figure}

Now we can analyze chargino contribution to the total decay width of $Z^0$-boson in a background electromagnetic field. In Standard Model the $Z^0$-boson decay width $\Gamma_Z$ into quarks and leptons in a strong electromagnetic field depends on the background field-strength parameter $\varkappa$: $\Gamma_Z =\Gamma_{\rm tot}(\varkappa)$   and it was calculated in our paper \cite{Kurilin-2009}. Relying on these studies it is possible to obtain branching ratios of $Z^0$-boson decay into charginos $B(Z^0\rightarrow\chi^+_i\chi^-_i)= \Gamma(Z^0\rightarrow\chi^+_i\chi^-_i)/ \Gamma_{\rm tot} (\varkappa)$ in a background electromagnetic field. Numerical calculations based on equation (\ref{Z-chi-1}) are presented in Fig. \ref{Branching-12}. We see that the processes of chargino production can be
significant in strong fields and it can give a sizable  contribution to the total decay width of $Z^0$-boson. In superstrong fields the branching ratio of decays into lighter charginos  $B(Z^0\rightarrow\chi^+_1\chi^-_1)$ is about $16\%$ while the branching ratio of decays into heavier ones $B(Z^0\rightarrow\chi^+_2\chi^-_2)$  can reach the values about $8\%$.

%%%%%%%%%%%%%%%%%%%%%%%%%% Fig2 %%%%%%%%%%%%%%%%%%%%%%%%%%%%%%%%%%%
%%%%%%%%%%%%%%%%%%%%%%%%%%%%%%%%%%%%%%%%%%%%%%%%%%%%%%%%%%%%%%%%%%%%

Thus summarizing the results obtained above we see that external electromagnetic
fields can change drastically the physics of quantum processes in a vacuum and serve as a catalyst for new phenomena and new physics. As the external-field strength increases, SUSY-decay modes of $Z^0$-boson could become observable and charginos $\chi^\pm_i$ predicted in the framework of MSSM could be detected. Although external-field strengths necessary for direct observation of the processes $Z^0\rightarrow\chi^+_i\chi^-_i$  are not yet available in experiments
there are promising expectations to realize similar extreme conditions in high intensity laser interactions \cite{ExtFields} or in physics of single crystals \cite{Baier1998}.

\end{document}